\documentstyle[11pt,newpasp,epsf,twoside]{article}
\def\edcomment#1{\iffalse\marginpar{\raggedright\sl#1\/}\else\relax\fi}
\reversemarginpar

\begin{document}
\title{Galaxy Disruption Caught in the Act}
 \author{A. M. Karick}
\affil{School of Physics, University of Melbourne, Victoria 3010, Australia}
\author{M. J. Drinkwater; M. West}
\affil{University of Queensland, Queensland 4027, Australia\\ Department of Physics and Astronomy, University of Hawaii, Hilo, HI 96720}
\author{M. D. Gregg; M. Hilker}
\affil{Department of Physics, University of California, Davis, CA 95616, USA\\ Sternwarte der Universit$\ddot{a}$t Bonn, Auf dem H$\ddot{u}$gel 71, 53121, Germany}

\begin{abstract}
Direct evidence of stellar material from galaxy disruption in the
intra-cluster medium (ICM) relies on challenging observations of
individual stars, planetary nebulae and diffuse optical light. Here we
show that the ultra-compact dwarf galaxies (UCDs) we have discovered in
the Fornax Cluster are a new and easy-to-measure probe of
disruption in the ICM. We present spectroscopic observations
supporting the hypothesis that the UCDs are the remnant nuclei of
tidally ``threshed'' dwarf galaxies. Deep optical imaging of the
cluster has revealed a 43-kpc long arc of tidal debris, flanking a
nucleated dwarf elliptical (dE,N) cluster member. We may be witnessing
galaxy threshing in action.
\end{abstract} 

\noindent{\bf 1. \hspace{1mm}Introduction}\\\\
Recently a number of spectacular streams of low surface brightness
material have been observed in galaxy clusters - evidence that galaxy
disruption is an ongoing and important process in galaxy
evolution. Material torn from galaxies is dispersed into the
intra-cluster medium, contributing up to 10-20\% of the cluster luminosity.

A small sample of ultra-compact dwarf galaxies were discovered during
an all-object Fornax Cluster Spectroscopic Survey (FCSS) using the
Anglo-Australian Telescope 2dF spectrograph (Drinkwater et al. 1999,
ApJ, 511, L97). These old stellar systems are 10 times more luminous
than the brightest globular clusters associated with the central
Fornax Cluster galaxy NGC 1399, with intrinsic sizes of $\sim$100 pc
and B-band magnitudes ranging from -13 to -11 (Karick et al. 2003,
MNRAS, 344, 188). Truly intergalactic, they are not associated with
any bright cluster galaxy. Our most favoured hypothesis for their
origin is that they are the remnant nuclei of infalling nucleated
dwarf elliptical (dE,Ns) galaxies which have been tidally ``threshed''
by large galaxies or by the cluster tidal field (Bekki et al 2001,
ApJ, 552, L105). \\
                              
\noindent{\bf 2. \hspace{1mm}Evidence for Galaxy Disruption}\\\\ The
internal velocity dispersions of the Fornax Cluster UCDs were measured
using the VLT and Keck Telescopes (Drinkwater et al. 2003, Natur, 423,
519). The velocity dispersions range from 24-37 km s$^{-1}$,
considerably higher than those of Galactic GCs. The UCDs lie well off
the globular cluster L $\propto$ $\sigma$$^{1.7}$ relation (Fig 1) in
a previously unoccupied region. The locations of the UCDs and the
dE,Ns on the diagram strongly support the galaxy threshing model.
A normal dE,N halo, accounting for $\sim$98\% of the dE,Ns luminosity, is
tidally stripped therefore reducing the total luminosity by a factor
of 100 but barely changing the central velocity dispersion. Over time
this disruption contributes diffuse stellar material to the ICM. The
surviving nuclei are dispersed into intra-cluster space or added to
the envelopes of brighter galaxies, where they can masquerade as
stars. This process may explain the observed high ``specific
frequency'' of GCs in central cluster galaxies.

We also obtained deep multicolour (u,g,r,i,z) imaging of the central 2
square degree region of the Fornax Cluster using the CTIO 4m MOSAIC
Telescope ($0.27''$pix$\approx$27 pc). This revealed a 43 kpc-long
plume of tidal debris 0.5 degrees from the cluster centre, flanking
the dE,N cluster member FCC 252 (Fig 2). The diffuse stellar material
has $\mu_g$=26.9 mag arcsec$^{-2}$ and $M_g$$\approx$ -14, comparable
to faint Local Group dwarf spheroidals (Mateo 1998, ARA\&A, 36, 435).
The proximity of the debris suggests we may be witnessing galaxy
threshing in action. The stellar populations of the UCDs, dE,Ns and
GCs associated with NGC 1399 will be determined from two-colour
diagrams. This will provide a further test of the galaxy threshing
hypothesis.  We require higher resolution observations of FCC 252 and
the debris to determine their possible tidal relationship.

\begin{figure}[h]
\plotone{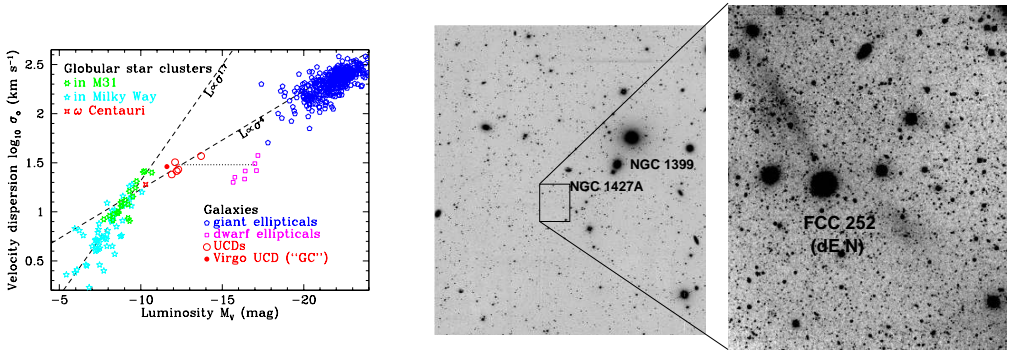}
\caption{Comparison of the internal dynamics of the UCDs with GCs and
normal galaxies. The horizontal line shows the predicted effect of galaxy
threshing. Figure 2:\hspace{2mm} Left: A 0.5$\deg$$\times$0.5$\deg$
region of the Fornax Cluster. NGC 1427A is part of an infalling
subcluster and shows evidence of interaction with the ICM. Right: An
$8'\times8'$region showing the curved arc of tidal debris, extending
from upper left to lower right through the dE,N, FCC 252.}
\end{figure}

\noindent {\bf Acknowledgements:} We would like to thank our collaborator R. Jurek

\end{document}